Title: The Possible Solution to the Two-Slit Electron Interference Pattern
Author: J.Zzimbe
Comments: PDF. The primary focus of this paper pertains to an analysis of the two-slit pattern. This comprises the initial twelve pages of this paper. However, since other aspects of quantum mechanics are addressed, the paper, in its entirety is twenty five pages in length.

To this day, the two-slit electron interference pattern remains shrouded in an inordinate mystery. It continues to defy a logical and rational explanation. This paper will postulate a new "characteristic" pertaining to the electron. If this characteristic is correct, the mystery of the two-slit pattern will be eliminated. Furthermore, the theory is testable. Fundamentally, the tests entail running the experiment in slightly different ways. If the theory is correct, we should see patterns that we've never previously witnessed.

I The Possible Solution to the Two-Slit Electron Interference Pattern

The two-slit electron interference pattern can be explained in common sense, logical terms without having to resort to any "nonsense" such as an electron propagating through two holes simultaneously. The central key to this mystery can be explained via the following concept.

> *The electron is only a particle. It is **not** a wave.*
> *Although an electron is not a wave, it does propagate*
> *with a **sinusoidal** motion.*

A mathematical description of this statement is as follows.

$$r = \sqrt{x^2 + y^2} \qquad (1)$$

In this equation, x and y denote the following

$$x = vt \qquad (2)$$

$$y = a\cos(\omega_o t + \phi) \qquad (3)$$

My *current* theoretical perspective is that the electron's sinusoidal motion is the result of some intrinsic dynamical factors of the electron, whereas the velocity is induced by the electron gun. Since, as of this writing, I am completely oblivious as to what intrinsic factors would induce this sinusoidal motion, my future theoretical perspective may change and cause me to adopt the position that the electron's velocity *and* sinusoidal motion are induced by some type of intrinsic dynamical factors. However, for the present, the two-slit electron interference pattern can (primarily) be analyzed via this concept. In order to most effectively outline the paths that the electrons are taking through the two-slit apparatus, it would be most judicious to outline what is transpiring when only one slit is open. (This may appear to be a redundancy, as there is no "mystery" pertaining to the singular slit. Although there is no "mystery", it is, nevertheless, important to outline what is transpiring in order to ensure comprehension when the two slit apparatus is analyzed.)

When only one slit is open, there are two alternative possibilities of what will transpire to the electron. One of these possibilities is fairly obvious, and, as has already been stated, there is no "mystery" pertaining to it. The electron simply propagates through the hole and becomes part of the emerging bright spot on the screen. (I trust that since this precept is reasonably simple, an illustrative diagram would be superfluous.) However, according to the precepts outlined in this paper, the electron propagates with a sinusoidal motion. Therefore, there is the possibility that when the electron is either at its "crest" or "trough", it can reflect from the *top* part of the hole, or from the *bottom* part of the hole. (The reader may object to the terminology utilized. The words crest or trough should only be utilized within the context of an actual wave, not a particle which propagates with a sinusoidal motion. Although there may be merit to an argument of this nature, the terms crest and trough are being utilized in order to convey certain precepts.) If this transpires, the electron will become "lost". To be specific, the "lost" electron will not become part of the emergent bright spot as it will not end up in the area *directly* behind the hole. The diagrams will serve to illustrate this principle.

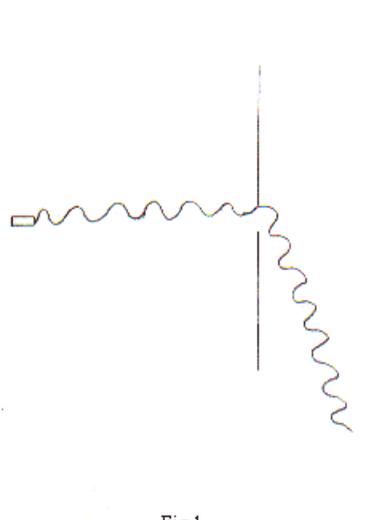

Fig.1

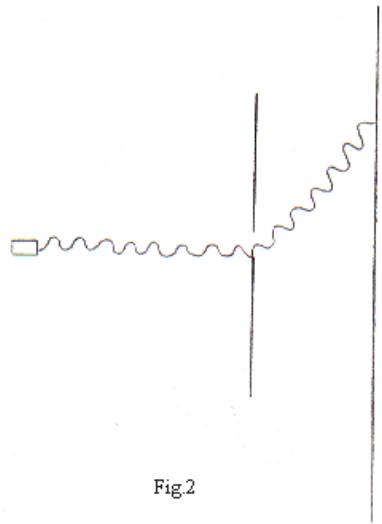

Fig.2

A question may naturally be raised. What factor will affect whether the electron will be at its crest, trough, or node when it reaches the hole? Precise details cannot be provided since, as has already been stated, I am oblivious as to *why* the electron propagates with a sinusoidal motion.[*] In equation three, it is the value of $f$ which will determine whether the electron is at its crest, trough, or node upon reaching the hole. The electron is in a state of constant oscillation. Therefore, each electron that leaves the gun will emerge with a different phase. One electron may emerge from the gun with its internal dynamics causing it to be at its node and propagating upwards. The next electron may emerge with its internal dynamics causing it to be at its crest and propagating downwards. There is a whole spectrum of phases with which the electron can potentially emerge from the gun. Therefore, the initial value of $f$ will be the primary factor which determines the final phase with which the electron reaches the hole.

An analysis of the two-slit electron interference pattern can now be implemented. The equations for the two-slit pattern remain essentially the same as the first three equations. However, there will be a difference amongst the following equations and equation number one. Equation one utilized the equation of a circle (as it pertained to the direction of propagation from the electron gun to the hole). The following equations will, instead, work within the parameters of Pythagoras' theorem. The reader may be wondering what the difference is since both equations are fundamentally similar ( $x^2 + y^2 = r^2$ or $x^2 + y^2 = z^2$ ). However, there is a dichotomy between the two equations as it pertains to the two-slit electron interference pattern. This dichotomy pertains to the *direction of propagation* of the electron. With the equation of a circle, the direction of propagation will be established by the equation characterizing its velocity. However, with Pythagoras' theorem, the direction of propagation will be along the hypotenuse. (The reader should keep in mind that right triangles can be established between the electron gun and the holes in the walls.) This is the direction of propagation required for the two-slit pattern.

In the following equations, numerical subscripts will be utilized as they will subsequently be modified in order to explain the full spectrum of the pattern. The electron's velocity is established via the following. (Once again, my current theoretical perspective dictates that the electron's velocity is extrinsic.)

$$x_1 = vt \qquad (4)$$

Its (intrinsic) sinusoidal motion can be characterized via the following.

$$y_1 = a\cos(w_o t + f) \qquad (5)$$

Once again, these two equations are combined to produce the following.

---

[*] Some readers may be of the following view. If I am incapable of providing precise details, this overall theory is worthless since science requires high precision. I beg to differ. If I can remove the *substantial* mystery of the two-slit electron interference pattern, then this theoretical framework is meritorious, even if it is lacking certain details. Furthermore, I am confident that in the future, I *will* ascertain why the electron propagates with a sinusoidal motion.

$$z_1 = \sqrt{x_1^2 + y_1^2} \qquad (6)$$

Although equation six constitutes the basic equation, it will subsequently modified (via modifications to equations four and five) in order to explain the full interference pattern.

We can now analyze the two-slit apparatus and explain the mystery of the interference pattern. Firstly, we shall establish certain "labels". The top hole (the hole closest to the top side of the paper) will be referred to as hole 1. The hole immediately beneath it will be referred to as hole 2. Since the explanation of the paths being traversed by the electrons can become potentially confusing, the exact paths will be described in steps. In essence, there are six primary paths that an electron can take when it propagates through the two-slit apparatus. Although there are incremental "variations" on these paths, there are six primary paths that an electron can take.

Two of those paths are as follows. An electron can propagate through the centre of hole 1 or hole 2 and become part of the pattern. To be specific, given the provision it propagates without interacting with the edges of the holes (i.e. the top or bottom parts of the holes), these would be two of the paths that it can take in order to become part of the final interference pattern. Regarding what has just been stated, it would be appropriate to emphasize the following. To state that an electron can potentially propagate through the centre of one of the two holes is somewhat simplistic. There is more to be stated as it pertains to the electron's propagation through the centre of a hole in order to explain the *full* interference spectrum that is produced. However, for explanatory purposes, it would be more judicious to subsequently present the requisite concepts in relation to an electron propagating through the centre of one of the two holes. For the time being, what has been stated will suffice. (It is deemed that the concept of an electron propagating through the centre of a hole is sufficiently simple that, once again, a diagram would be superfluous.)

The reader may be wondering, what factor will affect whether the electron propagates through hole 1 or propagates through hole 2? Prior to quantifying this, there will be a physical description of what will determine this. As already stated, the electron possesses an intrinsic oscillatory motion. If this motion is causing it to propagate upwards, *at the time of its emission from the gun*, it will propagate towards the top hole. If however, its intrinsic motion is causing it to propagate downwards, it will propagate towards the bottom hole.

Quantifying these statements will prove to be rather delicate. In fact, due to the method of quantification I have chosen, it would be most judicious to initially attempt to "assuage" the reader about its nature. Firstly, since I am postulating new physical principles, it is not unrealistic to propose new methods of quantification. Secondly, although the reader may be somewhat taken aback by what I am about to propose keep the following in mind. *No rules of mathematics are being violated.* In order to slightly elucidate upon the last statement, the following will be stated. If I were to state 3 + 4 x 5 = 35, regardless of what new physical theory I was postulating, it cannot be accepted as the rules of math are being violated. However, given the provision that the rules of math are not being violated, the grounds for opposition become less significant.

The method of quantifying whether or not the electron propagates through hole one or two is as follows. The electron's position in the y direction will be characterized by $\boldsymbol{c}$. If the electron were to propagate through the top hole, we will have the following equation.

$$y_2 = a\cos(\boldsymbol{w}_o t + \boldsymbol{f}) + \frac{\partial \boldsymbol{c}}{\partial t} \qquad (7)$$

If on the other hand, the electron propagates towards the second hole, the equation will be as follows.

$$y_3 = a\cos(\mathbf{w}_o t + \mathbf{f}) + \left(-\frac{\partial \mathbf{c}}{\partial t}\right) \qquad (8)$$

If the electron were to propagate through the second hole, this modification would cause equation six to change via the following.

$$z_2 = \sqrt{x_1^2 + y_3^2} \qquad (9)$$

The following question may be in the reader's mind. If it is $\mathbf{c}$ which will have bearing on whether the electron will propagate through hole 1 or hole 2, what will affect whether the electron is at its crest, node, or trough at the time it reaches the actual hole? A precept previously stated remains in effect in answering this question. Specifically, the *precise* value of $\mathbf{f}$ when the electron is emitted from the gun will affect whether it is at its crest, node, or trough at the time it reaches the actual hole.

The reader should be able to assess that this theoretical framework is being presented in stages. In establishing whether the electron will propagate through the first hole or the second hole, the value of $\mathbf{c}$ will affect this. When establishing if the electron will be at its crest, node, or trough upon reaching the hole, this will be established by the value of $\mathbf{f}$.

A few paragraphs ago, it was stated that it was insufficient to state that an electron propagated through the centre of one of the two holes. More needed to be stated in order to explain the full interference pattern. At this stage, an additional factor must be incorporated as it pertains to the propagation of an electron through the two-slit apparatus. This factor pertains to the polarization of the electron. When dealing with the two-slit pattern, there is a difference between an electron that propagates with a polarization of 90 degrees and one that propagates, as an example, with a polarization of 45 degrees. The difference would be that these two electrons would end up in *slightly* different areas on the screen. This is one of the factors which induces incremental differences which culminate in the full interference pattern witnessed in the course of experiments. Although it would be best to quantify the component of polarization within the equations, regrettably it is not judicious to proceed with its quantification for the following reason.

At this stage, I am oblivious as to whether the electron's polarization is the result of an intrinsic or extrinsic factor (or possibly some combination of the two). This factor would have bearing on the final equation. Consequently, it would be inappropriate to attempt to quantify the electron's polarization (within the parameters of the presented equations) until my analysis of what induces the electron's polarization is deepened. (The reader may feel that this analysis is becoming increasingly vague. However, in the end, the two-slit electron interference pattern will be explained in logical, common sense terms. In other words, the "mystery" of the pattern will be eliminated.)

The next two primary paths can be explained via the following. For the first path, let us assume that when the electron reaches hole 1, the value of $f$ is such that the electron will be at its crest. For the second path, let's assume that when the electron reaches hole 2, the value of $f$ will be such that the electron will be at its trough. If these values of $f$ were in effect, then an electron would reflect from the *top* of hole 1 (once again, the hole closest to the top side of the paper), or the *bottom* of hole 2. If this had transpired only when there was one hole, the electron would have become "lost" and *not* become part of the pattern. However, when there are two holes open, the electron can now become part of the interference pattern. If the electron reflects from the bottom of hole 2, it will reflect in an upwards direction and become part of the pattern. (See diagram.)

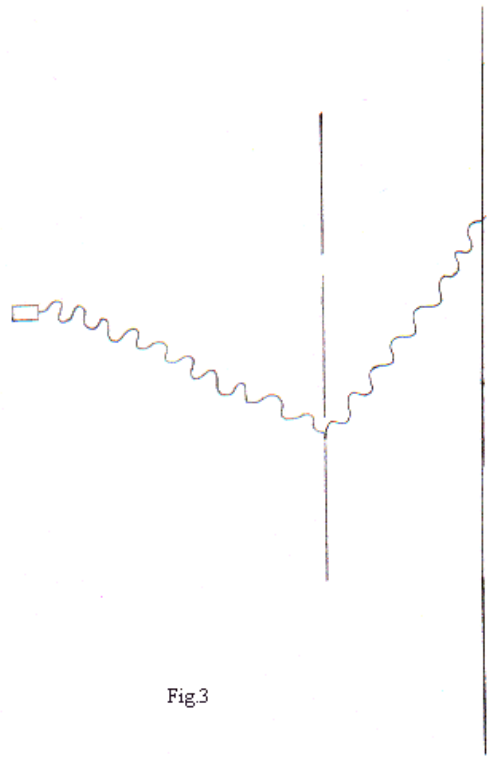

Fig.3

If this theory is fundamentally accurate, it will explain some very important aspects of the two-slit electron interference pattern. As was stated at the beginning of the paper, the interference pattern can be understood in logical, common sense terms. There is no need to resort to an electron simultaneously going through two holes, or an electron leaving as a particle, travelling as a wave, and finally landing as a particle. The path outlined in figure 3 explains the "illusion" of an electron going through two holes simultaneously. It *only* goes through one hole. However, it can end up in the area behind the other hole, thereby presenting the illusion of an electron propagating through two holes simultaneously. This would also explain the phenomenon of "constructive interference". (Constructive interference is in quotes as two forms of energy are *not* interacting which is the case with interference.) It goes without saying that all of these precepts apply if the electron reflects from the top of hole 1. The only difference being that the electron will reflect downwards and become part of the pattern by landing in the area behind hole 2.

It is reasonably important that something be stated in relation to diagram three. The diagram does not represent with complete accuracy the experimental results of the two-slit electron interference pattern. When an electron reflects from the top of hole 1 or the bottom of hole 2, the probability is that it will reflect to the area on the screen which is half way between the two holes. This, of course, is for the very simple reason that the experimental results of the pattern show that this is the area where there is the greatest concentration of electrons. However, in diagram three, this specific representation has not been shown for the reason already stated in the previous paragraph. Namely, the attempt to clearly illustrate the illusion of how a singular electron can seemingly propagate simultaneously through two holes. With this being stated, the reader should keep in mind that when an electron reflects from the top of hole 1 or the bottom of hole 2, there is a greater probability of it reflecting on to the area mid-way between the two holes (in accordance with the experimental results).

  Previously, when it was stated that electrons propagate through the centre of the holes without interacting with the edges, the following position was, subsequently, postulated. In order to effectively explain the full interference pattern, another dynamical element must be introduced (in that instance, polarization). In a similar way, when dealing with electrons that propagate by interacting with the edges of the holes, another dynamical element must be introduced in order to explain the full spectrum of the interference pattern. This other dynamical element pertains to the varying velocity of the electrons. It must be emphasized that the varying velocity does not entail an electron varying its velocity during the course of its propagation. It merely entails one electron experiencing a different velocity from another electron at the time it is emitted from the electron gun. Prior to expounding upon how electrons with different velocities would contribute to the spectrum of the interference pattern, it would be appropriate to answer what may be the reader's natural question. If electrons propagate with different velocities, what *induces* these different velocities?

  I have postulated the following. When electrons are still in the gun, there would be various quantum mechanical factors which would induce various spatial orientations *in relation* to other electrons (or possibly even protons). It would be this physical proximity to other electrons which would affect an electron's final velocity when it is emitted from the gun. Since the majority of the electrons which reflect from one of the two holes land in the area that is mid-way between the holes, we will accept that the majority of electrons propagate with one particular velocity. This entails that they are in a certain physical proximity to other electrons at the time of their emission from the electron gun. The electrons that propagate more slowly than these "average velocity electrons" are slightly further from other electrons at the time of their emission. Their propagation is slower as they experience a magnitude of repulsion which is smaller than the "average velocity electrons". Conversely, the electrons which propagate with a higher velocity are slightly closer to other electrons at the time of their emission thereby inducing a greater magnitude of repulsion.

  The specific details of how electrons with varying velocities contribute to the spectrum of the interference pattern can now be expounded upon. Firstly, when dealing with electrons which reflect from one of the edges of the holes, we will work within the parameters of the rudimentary physics law that the angle of incidence equals the angle of reflection. As was stated in the previous paragraph, the majority of electrons propagate with a velocity that induces them to land in the area that is mid-way between the two holes. If we were dealing with hole 1, what dynamical factors would cause an electron to land in area *below* the main peak? The electron would have to propagate at a slower velocity (in relation to an electron that would land in the area mid-way between the two holes) in order to reflect from an angle that would cause it land in an area below the main peak. It is postulated that the difference among the velocities of the electrons would be minimal. Therefore in the following equation, it will be stated that $0 < b < 1$.

$$x_2 = vbt \tag{10}$$

(It is a redundancy to continuously modify the final equation for "z" as it is presumed that the reader fully comprehends the nature of the modification without me having to write the actual equation.) Conversely, if we wanted to explain an electron that landed *above* the main peak, then the electron must propagate at a faster velocity than an "average electron" (i.e. an electron which was at the mid-way point of the spectrum) in order to ensure an appropriate angle of incidence/reflection. Consequently, the following equation would be applicable as it pertains to velocity.

$$x_3 = (v/b)t \tag{11}$$

These equations would suffice to explain the two-slit electron interference pattern. As an example, let's assume that an electron reflected from the bottom of hole two in such a manner that it landed above the main peak. The following would be in effect with the value of *f* causing the electron to be at its trough when it reaches the second hole.

$$z_3 = \sqrt{x_2^2 + y_3^2} \tag{12}$$

Now that the concept of varying velocities has been incorporated into the two-slit interference pattern, this concept must not only be applied to electrons that reflect from the edges of the holes, but also, to a minimal extent, to electrons which propagate through the centres of the holes. The precise nature of incorporating varying velocities to electrons which propagate through the centres of the holes is explained via the following.
Some electrons which propagate through the centre may propagate in such a way that when they reach the hole, they are *almost* (but not quite) at their crest (or trough). Consequently, they propagate through the hole without interacting with the edge. However, if the electron had reached the hole just a fraction of a second sooner (or later), then it would have been at its crest (or trough), and therefore reflected from the edge without propagating through the centre. Electrons which propagate with varying velocities (again, at the time they are emitted from the gun) will fall within these parameters and not propagate through the centres. Consequently, the varying velocities of electrons not only have bearing on electrons which reflect from the edges of the holes, but also electrons which propagate through the centres of the holes. (The reader may feel that there is a substantial flaw with what has just been proposed. Whatever statements are made pertaining to electrons should, to one extent or another, be applicable to photons as photons behave as electrons when dealing with the two-slit electron interference pattern. It would seem preposterous to give consideration to the possibility of various photons varying their velocities since the speed of light has been measured to several decimal places. However the reader should keep the following in mind. With the two-slit experiment, we are dealing with individual photons, not waves of light comprised of photons of different frequencies. Furthermore, the specific differences in the velocities would merely be fractions of a centimetre per second faster/slower than the "average velocity" photons. If one contrasts this with certain contemporary cosmological models which suggest that the reason for the homogeneity of the universe is that the speed of light in the early universe was $10^{38}$ times faster than it is today, then the concept of photons which are incrementally faster/slower than others does not seem so preposterous.)

The reader may feel there is a substantial flaw with equation twelve. Specifically, polarization by reflection has not been given due consideration. When an electron reflects from a certain type of surface, its angle of polarization will change. To be specific, if $\boldsymbol{q}_p$ denotes the polarization of the electron, then $\boldsymbol{q}_{p(\text{initial})} \neq \boldsymbol{q}_{p(\text{final})}$. Therefore, to incorporate this aspect into the equations for electrons which reflect from an edge, we will modify equation twelve with the following. (The reader may feel that this contradicts what was previously stated about polarization. Namely, polarization cannot be effectively quantified as I am oblivious as to whether the electron's polarization is the result of an extrinsic or intrinsic factor. However, notwithstanding this fact, in the current context we are now dealing with polarization by reflection. Since we know it is an extrinsic factor, it can be quantified.)

$$z_3 = \sqrt{x_2^2 + y_3^2} + \tan \boldsymbol{q}_p \qquad (13)$$

Finally, in order to *fully* explain the interference pattern, destructive interference must also be explained. To be specific, when the two-slit interference pattern is compared with the one-slit pattern, how is it that there are fewer electrons in certain places when a second hole is opened? Since there are a total of six primary paths that can be taken by the electrons, destructive interference can be explained via the final two paths. Those final two paths would be the *bottom* of hole 1, and the *top* of hole 2. If the electron were to reflect from the bottom of hole 1, it would reflect upwards, and become "lost". Specifically, it would not become part of the pattern. If the electron were to reflect from the top of hole 2, it would reflect downwards and become lost. (Since this concept has already been illustrated when dealing with the singular slit, a diagram is deemed superfluous). Although there would be different polarizations and velocities of the electrons, these variations are immaterial as the electron becomes lost and does not become part of the pattern. Therefore, there would not be any variations in the pattern as a result of varying polarizations and velocities. Equation nine would suffice to explain these final two paths with $y$ being either $y_2$ or $y_3$. Furthermore, the value of $\boldsymbol{f}$ must cause the electron to be at its crest when it reaches hole 2 or its trough when it reaches hole 1.

Although these six paths are sufficient to explain the electron two-slit interference pattern, it would be appropriate to mention one final concept. In order to illustrate this concept, let us assume that the experiment is conducted several different times. In each "run" of the experiment, one electron could take different paths. For example, let's take electron number 4269. In the first run, electron number 4269 may go through the centre of hole 1 and become part of the pattern. In the second run, #4269 may reflect from the top of hole 2, reflect downwards and become "lost" thereby preventing it from becoming part of the pattern. In the third run, it may reflect from the top of hole 1, become part of the pattern and present the "illusion" of the electron going through both holes simultaneously. Therefore, a particular electron can potentially take different paths and assume different polarizations and velocities during various "runs" of the experiment.

Prior to outlining experimental tests of this theory, let us make some superficial attempt to assess the viability of this theory. The attempt will only be superficial since every theory in physics requires *experimental* verification. However, as a starting point, it is viable to assess this particular theory by seeing if it makes intellectual sense and conforms to known observations.

On the point of being intellectually sensible, what has been proposed seems far more logical than claiming that an electron is simultaneously propagating through two holes. Furthermore, to state that an electron is a particle that propagates with a sinusoidal motion is far more rational than claiming that an electron "leaves as a particle, travels as a wave, and lands as a particle." It would not be feasible to compare every theory ever proposed to explain the interference pattern with the theory outlined in this paper. Suffice to state that this theory seems to effectively conform to the parameters of logical common sense.

Secondly, does this theory conform to known observations? In my view, the areas of observation that are effectively explained are as follows.

1. The interference pattern (both constructive and destructive) produced on the screen.
2. The "illusion" of an electron going through two holes simultaneously. Specifically, the electron propagates through one hole only, but has the potential to end up in the area behind the second hole.
3. The quantum mechanical penetration of a barrier such as is witnessed in the Josephson effect.

The third point outlined should be briefly dealt with. Contemporary quantum theory dictates that a particle cannot achieve quantum tunneling. Only a wave can tunnel. However, this would only be valid if a particle propagated with a completely linear motion. The reader should be well aware of the fact that this paper has emphatically stated, that although an electron is a particle, it does propagate with a sinusoidal motion. Therefore, a particle such as an electron would be able to achieve the quantum mechanical penetration of a barrier if it propagated with a sinusoidal motion.

Consequently, although we must achieve experimental verification of this theory, it can be viewed as perfectly viable in light of the fact that (a) it makes intellectual sense, and (b) it explains known observations. In physics, regardless of how sound a theory seems, experimental verification is required. The theory outlined is testable. Fundamentally, the test involves running the experiment in a slightly different manner. If the theory is correct, a slightly different pattern should emerge.

One test of this theory is as follows. Firstly, run the experiment in the "normal" way until the interference pattern emerges. For the test, close only one hole but leave the emitter in the same place (the spatial orientation of the emitter is reasonably important for this test). The test of this theory would require two changes. Firstly, close one hole (but, as already stated, leave the emitter in the same place). The second change would entail running the experiment for a time period *significantly* longer than the time required for the two-slit interference pattern to emerge. How much longer this would be, I cannot state. However, to be safe, let's state ten times longer than the time required for the two-slit pattern. If the theory of a particle propagating with a sinusoidal motion is correct, what should emerge is the following. After a substantial period of time, a spot should emerge behind the open hole. At the mid-way point, another spot should emerge. In the physical area above the mid-point, (assuming it is the second hole which is left open), successively dimmer points should emerge wherever the bright spots of the pattern are established during the course of the two-slit pattern. This would be the result of electrons (or photons) periodically reflecting from the edge of the open hole in such a manner that they end up in the area of the bright spots. Therefore, one test of this theory entails the following. Ensure the correct spatial orientation of the emitter, close one hole, and run the test for a period of time significantly longer than the two-slit pattern. There should be a spot behind the open hole and, with sufficient time, successively dimmer spots in the areas where bright spots are established during the course of the two-slit pattern. If the test is run in this manner, I am confident that this particular pattern will emerge.

    A second, somewhat similar test can be performed. However, I am not too confident about the predicted results emerging. Run the *two-slit* pattern for an inordinately long time. If the screen upon which the pattern is emerging is long enough, then at the "outer edges" of the pattern (or more specifically, the outer edges of the *screen* where the pattern emerges), dim spots may eventually emerge. These would be the particles which are reflecting from the edges in such a manner that they become "lost". The precise pattern of these spots would be difficult to predict. However, an approximation of the pattern is as follows. It goes without saying that these particles experience varying velocities and varying angles of polarization. Previously, these factors did not have to be given due consideration as they did not contribute to the two-slit pattern. However, for the purpose of this particular test, these factors become paramount. Presumably, a "minor" interference pattern (specifically one that is not as bright) may develop at the outer edges of the screen. The majority of particles would experience a certain average velocity. They, like the particles of the two-slit pattern, would land in a particular place at the outer edge of the screen. This would be the brightest spot in this area. Around this "bright" (relatively speaking) spot would be dimmer spots that are (somewhat) similar to the two-slit pattern. However, to see this pattern would require a truly inordinate time period. The reason being that for the two-slit pattern, there are particles from both holes contributing to the pattern. For the pattern just outlined, only particles from one hole are responsible for the pattern. Therefore, a truly inordinate time period would be required to see this pattern. Furthermore, the length of time required for this is not the only difficulty in seeing this pattern. The angle of incidence of the particles may be such that they do not land on the screen where the pattern is emerging. However, as already stated, a sufficiently long screen *may* circumvent this problem.

    A third test is as follows. For the conduction of this next experiment, the experiment must be performed with photons and there must be a material which *completely* absorbs photons. Let us assume that black felt will suffice. Ensure that the top of hole 1 and the bottom of hole 2 is covered with felt so that photons do not reflect from these areas. If the photons are completely absorbed, constructive interference will be eliminated and the peaks in the pattern should vanish. What should appear are two spots of light behind the holes. If experiment two was successful, a slightly different test of test three would be as follows. Run this third experiment for a substantial period of time and two very bright spots should appear behind the holes and some type of interference pattern should appear at the outermost edges of the screen as in the second experiment.

    If the third test is successful (its complete success is contingent upon the photons being completely absorbed by the material) then an interesting fourth test can be run. Cover *either* the top of hole 1 with felt, *or* the bottom of hole 2. Let's assume the bottom of hole 2 is covered. The following pattern should emerge. Behind hole 1 there should be a spot of light. In the area that is mid-way between the two holes, although there will be a peak, it should be half as bright as the peak produced when both slits are open. Furthermore, in the general area behind hole 2, an

interference pattern should emerge. This is for the simple reason that there are photons reflecting downwards from the top of hole 1, but no photons reflecting upwards from the bottom of hole 2.

Utilizing this similar type of premise (covering a particular area with black felt) the paths of the photons from the bottom of hole 1 or the top of hole 2 can be verified. In one type of experiment, cover both of these areas and run the experiment for an inordinate period of time. Given the provision that the felt completely absorbs the photons, an interference pattern should *not* emerge at the outer edges of the screen. Furthermore, cover *one* of these areas (let's say the bottom of hole 1) and run the experiment for an inordinate time period. What should emerge, is a very bright interference pattern behind the holes, some type of interference pattern at the lowermost edges of the screen, and absolutely nothing at the uppermost edges of the screen. The converse would be true if the top of hole 2 were covered.

Finally, if one desires it, cover both edges of both holes and run the experiment for an inordinate period of time. One should *only* witness two very bright spots behind the holes. In this experiment, there would be no constructive or destructive interference since the areas responsible for interference are now absorbing photons and not reflecting them.

Although we must wait for the experimental results of these tests, we can, nevertheless, assess the impact of this theory on the Heisenberg uncertainty principle. As the reader is well aware, the uncertainty principle states $\Delta x \Delta p \geq \frac{1}{2}\hbar$. The Heisenberg uncertainty principle, in essence, proceeds from the following basis. It is impossible to simultaneously ascertain the position and momentum of a particle as experiments reveal that a particle possesses *both* wave and particle properties. If we proceed from that point, then a substantial wave train is required to ascertain momentum thereby preventing knowledge of the particle's position. Conversely, if we ascertain position, we have no wave train thereby preventing knowledge of the particle's momentum.

However, since a theory has now been presented which shows that an electron is *not* a wave, is this principle true? If the electron were a wave (as opposed to merely propagating with a sinusoidal motion) what would be a pre-requisite to ascertaining momentum? A pre-requisite would be a substantial wave train. This would negate the ability to ascertain position. However, an electron that was only a particle which propagated with a sinusoidal motion, would not have this "deleterious effect" associated with it. Since $P = mv$ we do not require a wave train in order to ascertain momentum. We merely require its mass and velocity. A particle propagating with a sinusoidal motion would provide this. Therefore, we *can* ascertain both the position and momentum of a particle with an arbitrary degree of accuracy. This, of course, is in direct contradiction to the Heisenberg uncertainty principle.

Many readers may be of the following view. If the Heisenberg uncertainty principle is overturned, this would precipitate a substantial crisis within physics. However, I have developed alternative quantum theories *which are consistent with the experimental facts*. Furthermore, this alternative framework can answer questions which the Standard Model cannot answer. However, this theoretical framework will not be presented unless experimental verification (as it pertains to the two-slit pattern) is procured. This is for the simple reason that the alternative model is of a highly radical nature. The *only* possible way the scientific community would consider it, is if the Heisenberg uncertainty principle is overturned. This would *compel* the quest for new quantum theories. For those of you who may still be reluctant to proceed with these experiments, the following should be pointed out. In the past, Nobel prizes have been awarded for the experimental verification of theory. Although I am not suggesting that a Nobel prize would necessarily be awarded for the verification of this theory, the point is that these types of papers *are* important. If a paper verified these experiments, it may facilitate the attainment of a professorship for a post doctoral fellow.

Section I constitutes the primary "focus" of this paper. The rest of the paper pertains to other aspects of quantum mechanics and is of a far more abstract nature. Therefore, you are strongly advised *not* to read any further. However, there is one other aspect of this paper which the reader will probably find intriguing. As preposterous as it may sound, it *is* possible to ascertain whether Schrodinger's cat is *either* alive, *or* dead, but *not* in a superposition of states between life and death. This analysis will be found in section IV of this paper on page sixteen. Since this archive is unrefereed, some readers may find it to be nonsense. Therefore, the following course of action is suggested. In section IV, you will find "instructions" of where to read quickly and where to read more carefully. To be specific, read the initial few pages rather quickly just to gain a feel for the concepts being outlined. When it comes to the specific analysis of Schrodinger's cat, there will be instructions to read more carefully in order to ensure an appropriate level of comprehension.

II  Two philosophical precepts

The reader may be of the following view. Even if this theory does eventually culminate in the Heisenberg uncertainty principle being overturned, certain aspects of quantum mechanics would still remain firmly established. Specifically a probabilistic interpretation of quantum mechanics would still be entrenched as I myself have admitted that a designated electron (electron number "x") could potentially traverse different paths during various "runs" of the electron two-slit interference pattern. On this point (a probabilistic interpretation of quantum mechanics), how would Einstein feel about what has been proposed? At first glance, one may be inclined to claim that he would embrace this material as a result of his vehement opposition to the uncertainty principle. Namely, he could never accept that "God would place dice with the cosmos." However, in accordance with what has been outlined *thus far*, "God is still playing dice with the cosmos." Therefore, although Einstein *may* be somewhat amenable to what is being proposed (since the validity of the Heisenberg uncertainty principle is effectively being brought into question), in all probability he would not fully embrace what has (thus far) been outlined.

I too would tend to be opposed to a probabilistic interpretation of quantum mechanics. This is not because of how "God would play dice with the cosmos," but because of two philosophical precepts. It is a result of these principles that the *future* course of a particle can be predicted with absolute certainty (even if only in principle). The next section will outline the precise nature of these philosophical precepts.

It would be appropriate to state that when the reader is reading the first philosophical precept, it may seem nonsensical and irrelevant to an attempt to overturn a probabilistic interpretation of quantum mechanics or the Heisenberg uncertainty principle in general. However, it *does* have bearing on the probabilistic interpretation of quantum mechanics. Hopefully, this will become evident in due course.

In certain aspects of science, there seems to exist a certain fundamental belief that there are some questions which can never be answered. Some examples of this would be as follows. Within cosmology a question has been posed, "If the big bang was the moment of creation, what transpired prior to that?". The standard response seems to be that it is inherently unknowable as to what transpired prior to the big bang. Another example of a question which can seemingly never be answered is in *The Feynman Lectures on Physics* (Volume III, Chapter 1, page 10), when Feynman is discussing the two-slit electron interference pattern. He states the following. "One might still like to ask: 'How does it work? What is the machinery behind the law?' No one has found any machinery behind the law. No one can 'explain' any more than we have just 'explained.' No one will give you any deeper representation of the situation. We have no ideas about a more basic mechanism from which these results can be deduced." Even in chemistry, similar tenets are seemingly being expressed. When studying bonding, there is something known as a resonance structure (this should not be confused with the concept of resonance in high energy particle physics). A resonance structure can more or less be explained along the following lines. With certain molecules, chemists cannot develop an electron-dot formula that conforms to the spectral analysis produced by that molecule. Therefore, they draw at least two electron-dot diagrams and state that the molecule is a hybrid representation of these imaginary diagrams. In short, they do not have a conclusive representation that conforms to the spectrum produced by the molecule.

In all of these examples, there seems to be a common precept which can be expressed as follows. "It's a mystery which we will never be able to comprehend or explain." As far as I am concerned, a precept of this nature is not even incrementally acceptable. This is science, not the Roman Catholic Church. Although I will not agree with it (as I have no religious beliefs of my own) I will respect the statement, "It's a mystery," if you are making the statement as a catholic. Specifically, there is God the father, God the holy spirit, and God the son. How can God be in three separate forms and yet all be part of a singular God? Catholic theology dictates that it is a mystery which will never be understood. That's fine for religion. However, "It's a mystery which will never be understood," cannot be expressed within the parameters of science. Merely because an effective explanation is completely beyond our current capability, that doesn't necessarily mean that it will never be explained. As an example, I read an old geology book which delved into historical geology and briefly dealt with the extinction of the dinosaurs. Although it explained some theories about what may have caused their extinction, the book went on to state that we shall probably never know the complete solution to this mystery. However, theorists persevered. As far as I know, paleontologists world wide acknowledge that it was a meteor crashing into the earth which was responsible for the extinction of the dinosaurs.

If some phenomenon of nature cannot be explained, a physicist has no right to adopt the position of, "It's a mystery." Various phenomena may appear to be incomprehensible. However, regardless of how incomprehensible a designated phenomenon may seem, we must persevere with our attempts to probe and understand the phenomenon until an effective theoretical framework is developed. With time, our efforts will probably by fruitful.

The first philosophical precept that is germane to this paper is as follows.

> *Within the parameters of science, it is completely unacceptable to adopt a position which is reflective of, "It's a mystery which we will never be able to comprehend."*

The second philosophical precept is as follows.

> *The outcome of an event can be predicted with **absolute certainty** if an inordinate amount of data pertaining to the variables which will affect that outcome is available to us.*

It must be emphasized that in various instances this would only be true in principle. (The reader may already feel that this precept is invalid as the Heisenberg uncertainty principle states that position and momentum cannot be simultaneously ascertained. However, part I of this paper should have caused the validity of that principle to be seriously questioned.) Before proceeding to outline how this has bearing on quantum mechanics, it would be best to provide an illustration of this philosophical precept via a macroscopic "event".

Various sporting events have odds established on their outcomes in order to facilitate the placement of wagers. In other words, there is a probabilistic interpretation of their final outcome. However, we can, *in principle*, predict their outcomes with *absolute certainty* if we were in possession of an inordinate amount of data. As an example, let's take a boxing match. In principle, we do not need to lay 5 to 3, 10 to 1, or 7 to 4 odds on a particular fighter. If we had precise data pertaining to their strength, stamina, speed, conditioning, strategies, psychological makeup (will they endure no matter what, or go down after some pain?), state of mind prior to the fight (do they take their opponent seriously, or are they complacent?), previous life experiences that may have bearing on the fight, etc., etc., etc., then, upon a highly complex calculation, we could predict with absolute certainty who would win the fight, in what round, and the manner in which victory was achieved (knockout, or a decision). Obviously, to obtain such an inordinate amount of data and implement the requisite calculation is impossible in practical terms. However, in principle, it could be done.

We are now ready to state how these two philosophical precepts (specifically, don't state it's a mystery, and an inordinate amount of data will enable us to predict with absolute precision the outcome of an event) have bearing on the probabilistic interpretation of quantum mechanics.

It has been previously stated that a designated electron can take completely different paths during various "runs" of the experiment. We have no idea which paths may be taken during the course of a particular "run". Therefore, from all outward appearances the probabilistic interpretation of quantum mechanics is as strongly entrenched as ever. However, the two philosophical precepts outlined in part II of this paper, should bring that concept into question. We cannot state, "It's a mystery," when trying to assess which of the six primary paths an electron will take. *At this stage of scientific development* we are oblivious as to which path an electron may take. Therefore, the development of theory must continue until we develop an effective theoretical framework that will enable us to comprehend (even if it's only in principle) the variables that will cause an electron to take one of the six paths as well as its velocity and polarization. Consequently, if a physicist wants to claim that this is a mystery which will never be solved, I will only respect that statement if you are planning on giving up physics and aspiring to become the next Pope of the Roman Catholic Church. Otherwise, don't state, "It's a mystery," and expect me to accept that answer from a scientist. One day we will grasp the variables that affect an electron's trajectory, velocity, and polarization.

At this stage of the paper, there is the possibility that Einstein would be more receptive to what is being proposed. The validity of the Heisenberg uncertainty principle and other elements of quantum mechanics are being brought into question. Furthermore, from Einstein's perspective, God is no longer playing dice with the cosmos.

III The ramifications for certain elements of quantum mechanics

Working within the parameters of the theories and principles thus far outlined, *other* elements of quantum mechanics should now be dealt with.

Another theoretical aspect of quantum mechanics that would have to be questioned would entail Feynman path integrals. The path integral approach dictates that we must take into consideration every possible path that an electron can take and integrate in order to ascertain the most probable path. However, according to the solution to the two-slit interference pattern outlined in this paper, an electron which will propagate through a hole will only take one of six primary paths. Consequently, we do not have to take into consideration "every possible path" that an electron can take. It is a question of developing a theoretical model of the factors which will affect an electron's path and then gaining a sufficient amount of data which will enable us to predict its precise path, even if this is only possible in principle.

The philosophical precepts outlined in section II of this paper would have bearing on a Gedanken from quantum mechanics. Namely, Schrodinger's cat. The two philosophical precepts outlined indicate that we do not have to look in the box (and subsequently collapse the cat's wave function) to ascertain if the cat is alive or dead. Contemporary tenets dictate that we cannot, even in principle know whether a radioactive material will, or will not decay within a given time $\tau$. Therefore what are physicists saying? "It's a mystery." As far as I am concerned, if we cannot predict, even in principle, whether a radioactive material will decay or not, then our knowledge and understanding of radioactivity is incomplete. If it is of interest to know the exact moment when a radioactive material will decay, then we must develop our theoretical model of radioactivity until we comprehend the variables that will affect radioactive decay. In a similar way to the two-slit electron interference pattern, given an inordinate amount of data, we can, in principle know *if* the material will decay and the exact moment that it will do so. Therefore, working within the parameters of the philosophical precepts outlined in section II, Schrodinger's cat is not in a superposition of states between life and death until we open the box to look inside. It is *either* alive *or* dead but not in a superposition of states. This is the result of being able to predict the exact moment the material decays (if it decays at all) once we increase our knowledge of the theoretical model of radioactivity and gain an inordinate amount of data as it pertains to decay once it is time to make a prediction.

IV The theoretical possibility of measuring a quantum entity without collapsing its wave function

*Read quickly*

Before proceeding, it would only be fair to state that the title of this section may be of a somewhat misleading nature. There will be no tangible or concrete method outlined that would enable us to achieve what is stated in the title as of "tomorrow". Clarification of what *will* be achieved in this section is as follows.

It seems that it is absolutely impossible to measure a quantum entity without simultaneously collapsing its wave function. Let's revisit Feynman and see what he has to say about the issue.

The reference is the same as above except this is from page nine. "In our experiment we find that it is impossible to arrange the light in such a way that one can tell which hole the electron went through, and at the same time not disturb the pattern. It was suggested by Heisenberg that the then new laws of nature could only be consistent if there were some basic limitation on our experimental capabilities not previously recognized. He proposed, as a general principle, his *uncertainty principle*, which we can state in terms of our experiment as follows: 'It is impossible to design an apparatus to determine which hole the electron passes through, that will not at the same time disturb the electrons enough to destroy the interference pattern.' If an apparatus is capable of determining which hole the electron goes through, it *cannot* be so delicate that it does not disturb the pattern in an essential way. No one has ever found (or even thought of) a way around the uncertainty principle." What is particularly important for this section are the words in brackets. Namely, "or even thought of."

    Merely because no one has developed a method of measuring an electron without simultaneously collapsing its wave function as of yet, we must keep in mind that it doesn't necessarily mean there isn't a method. Feynman himself seems to recognize the possibility of measuring a quantum entity without destroying the interference pattern. He writes, "But if a way to 'beat' the uncertainty principle were ever discovered,...." I have not found a concrete method of how to "beat" it as of yet. However, I have *thought* of a way of circumventing it. I will outline a new method of taking a measurement which has not been given appropriate consideration when dealing with the quantum level. With continued development and time (many years) it may eventually culminate in the development of a technology which will enable us to accomplish what Feynman states cannot be accomplished. In essence, I will be challenging the alleged impossibility of measuring a quantum entity without collapsing its wave function via a new method of quantum measurement.

    If I am to ensure the lucidity of the new method of measurement that I am advocating, it is necessary that I first draw an analogy with submarine warfare. The reader is presumably cognizant of the fact that electromagnetic waves do not propagate in deep water. Since a submarine cannot utilize radar to locate an enemy submarine, it must rely on sound for detection purposes. Therefore, one method at a submariner's disposal of detecting an opposing submarine is to utilize its sonar to emit sound waves. If the waves reflect off another submarine, the waves will return to the detector and provide the location of the opposing submarine. This is somewhat similar to the measurements we strive to make on a quantum level. Namely, a source emits photons and, upon encountering an electron, those photons are reflected back to a detector in order to provide a measurement. Unfortunately, these same photons result in collapsing the electron's wave function. It is also "unacceptable" for a submarine to emit such a substantial amount of energy in order to detect an opposing submarine. On a tactical level, this will give away the submarine's own position and be contrary to the advantage offered by submarines (near invisibility). Therefore, the vast majority of the time (almost always), a submarine employs a second system for detecting enemy submarines. Namely, *passive detection*. What this specifically entails is the following. A submarine is outfitted with over 1000 microphones positioned at a myriad of different angles to detect any noise emissions from another submarine. The passive listening system is highly efficient as it has been known that sounds from *within* an opposing submarine have been heard by American submarines following Soviet submarines (at close range).

    Consequently, at a macroscopic level, there are two methods of rendering a measurement. The first method is an active system. Specifically, emit some form of energy and detect any

potential reflections. The second is a passive detection system. Namely, one's detectors are designed to detect an emission of some kind from what it is that one is trying to measure.

On a quantum level, we have only employed active detection systems. We have directed photons at the electrons and attempted to detect the reflected photons. Thus far, nobody has given consideration to the possibility of some sort of passive detection system for the quantum mechanical world. Obviously, an effective passive detection system for the quantum level would require new concepts and highly advanced technology. (The following should be pointed out. In regards to what has just been stated, the reader may be of the following view. An accelerating electron will emit photons. Perhaps this would fall within the requisite parameters of the passive detection system that is being advocated. It does not conform to the acceptable parameters of passive detection as once an electron emits a photon the emitted photon will cause the electron's momentum and position to change. Besides, accelerating an electron through the two-slit apparatus would probably cause the interference pattern to be destroyed.)

Before proceeding any further, it is imperative that something should be emphasized as it pertains to passive detection. Macroscopically, there are two passive detection systems that I am cognizant of. One pertains to submarine warfare and the other pertains to air and land warfare. The latter entails an infrared detection system which detects thermal radiation. In both of these cases, "the entity" which is being located emits some type of energy. However, this is only one of two types of passive detection. There is a second element to passive detection. The object which is being located can "make its presence felt" in one respect or another. Let's utilize another analogy from the military to elucidate on this principle. Stealth aircraft cannot be detected via an active detection system such as high frequency radar. However, that doesn't mean the aircraft is undetectable. As it moves through the air, it creates a substantial amount of turbulence, thereby "making its presence felt." This turbulence can be detected and reveal the location of the aircraft. This type of detection (making its presence felt) *already* exists within science. No type of weapons platform is being referred to, but rather *science* itself. Specifically, the detection of planets around other stars within our galaxy. How are these planets detected? Through passive detection and how the planet "makes its presence felt." Light is *not* reflecting from the planet and entering our telescopes here on earth. Instead, as the planet orbits around a star, its gravitational effect induces a slight "pull" (loosely speaking) on the star. It is this "perturbation" in the star's light that tells us that a planet is there. Therefore, passive detection can manifest itself in two different ways. The first is via "the entity" emitting something, the second is by "the entity" "making its presence felt" in one respect or another.

A very simple idea for passively detecting an electron without collapsing its wave function is as follows. As the electron passes through one of the holes, there will be some *incremental* change in the "system" (the wall with the holes in it). Specifically, the coulomb repulsion will be "in effect". When the electron comes "within range" of the valence electrons in the edge of one of the holes, this will induce some infinitesimally small change in the electrons which will produce and equally small change in the system which may be detectable via highly advanced technology. (Since the coulomb repulsion is inversely proportional to $r^2$, the electron may not be close enough to induce the requisite change. Therefore, it may be necessary to introduce some change in the wall/holes to facilitate this procedure.)

The reader may feel that this concept is so rudimentary that it is fundamentally useless towards the attainment of the stated goal. However, rudimentary concepts may, in the long run, culminate in the development of highly effective technology. An example to illustrate this principle is as follows.

The basic mechanism for a heat seeking missile was first developed in the 1950's by a scientist. However, at the time of the inception of the idea, it was completely useless as a weapon as it had to be inches away from the opposing plane's engine. However, what was important was that a primitive, yet effective concept was created. With continued research and development, the basic concept was developed to the point where it could be utilized for the purpose it was originally designed for. Today's missiles are so sensitive, they do not have to lock on to the heat from another engine. They can lock on to the heat generated by the friction of the wing with the air. Consequently, what initially began as a useless and simplistic idea, was eventually transformed into a highly potent, reliable system. Therefore, given the provision that an effective concept is developed, it is fundamentally irrelevant how simplistic that concept is towards achieving the requisite goal in the short term. Given the provision that it is effective, there is the strong possibility that continued research and development will eventually culminate in the attainment of the requisite goal.

*Read with greater care*

The main aspects of section IV of this paper were once shown to a physics professor. Of his many comments, one of them was as follows. "We <u>know</u> the quantum 'world' behaves differently from the classical one, so the existence of passive sonar shows nothing." I beg to differ. I can illustrate, via a concrete, tangible example that passive detection *does* have value within the framework of quantum mechanics. A vital Gedanken of quantum mechanics is rendered "null and void" as a direct result of passive detection.

Let's revisit Schrodinger's cat. Independently of the philosophical precepts outlined in section II, it seems certain that the cat is in a superposition of states between life and death until we open the box, have photons reflect off of the cat (in order to look inside), and thereby collapse the wave function of the cat. However, even without section II's philosophical precepts, this idea is completely wrong. It can be shown, that *without opening the box at all to look inside* (and subsequently collapsing the cat's wave function), we can ascertain, *with absolute certainty* that the cat is *either* alive *or* dead, but *not* in a superposition of states between life and death. All of this can be accomplished via passive detection.

Firstly, we shall have to introduce a slight modification to this Gedanken. (The reader may feel that this is "cheating". However, the modification will be minimal and the "essence" of Schrodinger's cat will remain in effect.) Even if the hammer did not smash the jar and release the poison, the cat would still die. Namely, it would suffocate to death due to a lack of oxygen and the buildup of carbon dioxide. Therefore, in order to conduct this thought experiment, we must ensure that the cat can breathe by introducing oxygen and removing carbon dioxide. Therefore, within the box we shall incorporate a circulation system via pumps that are equipped with detectors. When the detector senses carbon dioxide, it will pump the carbon dioxide out. The second pump will monitor the oxygen levels and, when they become too low, oxygen will be pumped into the box. However, independent of these two connections with the outside world, the box *must* remain closed so that no photons can reflect from the cat and collapse its wave function. To take this one step further, we can even put this contraption in a sealed room without any human observers so that we don't even see the box. As long as the detectors can be monitored (let's say by a computer) the experiment will work.

The cat and the diabolical device are placed in the box, the box is sealed, and the box is placed in a closed room that is devoid of any human observers. The humans are in another room, monitoring the computer which monitors the detectors. We now wait for a time τ. Since we can't look inside the box (or even see the box for that matter) how do we know if the cat is alive, or dead? Through passive detection. If the cat is alive, the carbon dioxide content of the box should be continually increasing. If the cat is dead, there will be no carbon dioxide as the pump has removed it from the box. Therefore, by assessing the carbon dioxide in the box (or lack thereof), we can ascertain that the cat is either alive or dead but not in a superposition of states between life and death. Most importantly of all, this is accomplished without collapsing the cat's wave function. It should be pointed out, that the detection of carbon dioxide is only one of two methods of implementing passive detection. The second pertains to how the cat "makes its presence felt." Instead of detecting the carbon dioxide, we can alternatively detect the oxygen in the box. As long as the oxygen is continuously being depleted, the cat is alive. If the oxygen level remains at the same level, then the cat is dead. I am cognizant of the fact that in both of these cases, the gas in the box is being monitored. Therefore, the reader may feel that the same basic method is being employed. However, there is a slight difference between the two. When the carbon dioxide is being detected, an emission from the cat is being assessed. When the oxygen is being detected, we are assessing how the cat is "making its presence felt."

Schrodinger's cat is rendered null and void. There is no longer a paradox, and this Gedanken can be eliminated from the framework of quantum mechanics. Consequently, the elimination of this alleged paradox illustrates that passive detection does have value within quantum mechanics.

*This completes the analysis of Schrodinger's cat.*

The reader may raise objections along the following lines. Firstly, this proposal (the possibility of passive detection on a quantum level in order to measure a quantum entity without collapsing its wave function) cannot be taken seriously as a *tangible, reliable* idea for the development of a technological system that would enable us to detect electrons on a passive level has not been presented. Secondly, even if such an idea were presented, it would be ludicrous to give consideration to such a precept as the development of technology of this nature is clearly impossible. Both issues will be addressed separately.

Regarding the first objection, the reader should keep in mind as to what was stated at the beginning of this section. It was clearly stated that no concrete method would be proposed. Only a broad, abstract method of taking a measurement would be outlined. The primary purpose of this section is to merely challenge the concept of the alleged impossibility of measuring a quantum entity without collapsing its wave function. Only the most rudimentary ideas will be presented in order to accomplish the stated goal.

At this stage, the reader may be adhering to the second objection. Namely, that the development of technology of this nature is categorically impossible. A view of this nature is terribly short sighted and narrow minded. The rest of section IV will be spent addressing this issue (namely, "the development of a certain type of technology is impossible").

We have no idea what the future holds. What new scientific principles and/or ideas may be developed that would permit the development of new technology that we would find astounding? Keep the following principle in mind. *Merely because something is beyond our present day scientific/technological capabilities, that doesn't necessarily mean that it will never become a reality.* Let's illustrate this principle via a hypothetical scenario. For this hypothetical scenario, we will return to the 1500's.

Let's assume that an individual were to approach the top scientists of the 1500's and state the following. "In my opinion, the following technologies will one day be possible. Any sort of an event, such as a jousting event, will be able to be seen by people thousands of miles away, at the instant it is occurring. I believe it will be possible to record the event, transmit the event in the air via special signals, and have someone thousands of miles away simultaneously witness the event when these special signals are received via a special device. I have even developed a name for this technology. I have called it, television. Furthermore, I believe it will one day be possible to send men from the earth to the moon in special ships and have them safely returned to the earth." What would the reaction of the top scientists of the 1500's have been to these proposals? Obviously they would have laughed at and ridiculed the entire notion. The technological possibility of television or landing a man on the moon was *completely* beyond the scientific/technological capabilities of the 1500's. There was no knowledge at the time which would have enabled them to even begin the construction of such technologies. Therefore, they would have dismissed the proposal as a fantasy which was completely unrealistic and not worthy of even incremental consideration. They would have insisted that not in 100 years, 1000 years, or even 1,000,000 years will such things be possible. They will *never* become a reality.

Let's take this hypothetical scenario a little further. Let's say that we could bring Newton to our present day world. Wouldn't he (as well as any other person from that age) be absolutely flabbergasted by the technologies of our world? Airplanes, television, radar, cars, radio, etc. In fact if in Newton's time someone had said, "One day doors will be able to open once you are within proximity to them without anyone or anything physically touching it," what would he have said? "By my first law of motion, that is impossible. Such an occurrence would violate fundamental laws of physics. It will *never* become a reality." He had no way of foreseeing the theorizing of the electromagnetic force or the development of the photoelectric effect. Therefore, if Newton were here today, he may very well faint when he first walked into a grocery store, and saw the door open without anyone touching it. In a similar way, we ourselves would be flabbergasted by technology if we could "magically" transport ourselves sufficiently far into the future.

It is possible to draw another analogy to address the issue of, "it's impossible to develop such technology", that is, in essence, realistic. The hypothetical components of this analogy will be minimal. When the theory of Bose-Einstein condensation was developed, it was clearly known that only bosons would be able to achieve BEC. Clearly, the science shows that fermions would be completely incapable of achieving BEC. The hypothetical component of this analogy is as follows. When Bose-Einstein condensation was proposed, there could have been individuals who adopted the following position. "It is conclusively and unequivocally impossible for fermions to achieve BEC. In this instance, it is not a question of seeing what new ideas may be developed in the future or what scientific discoveries may render BEC possible for fermions as a result of the following. The basis of technology is science. The *science* of BEC dictates that it is impossible for fermions to condense into the same quantum state. Therefore, regardless of what new technologies may be developed, or what new ideas may be proposed, fermions will *never* achieve BEC. It is simply impossible for such a thing to be accomplished." We will now return to the realistic components of this analogy. A few decades elapse and BCS is proposed to explain superconductivity. This paper is epochal in scope for various reasons. One element of its epochal nature entails the proposal of fermion coupling in order to achieve a composite boson. Eventually, the reality of fermion coupling is established on firm experimental grounds. The possibility of fermion coupling is then extended to BEC and we eventually realize that it *is* possible for fermions to achieve BEC if they couple with another fermion to establish a composite boson. Therefore, we see that our hypothetical individuals who adopted the position of, "...fermions will *never* achieve BEC", were wrong. We simply don't know what the future of science holds or what new ideas will be proposed. Hopefully, the reader will recognize the reality of, *merely because something is completely beyond our present day scientific/technological capabilities, that doesn't necessarily mean that it will never become a reality*.

Section IV can be summarized as follows. Feynman insists that an apparatus cannot be devised that will tell us which hole the electron went through without simultaneously destroying the interference pattern. However, he acknowledges that if a way to "beat" this idea were developed, things would change. Macroscopically, both active and passive detection systems exist. Quantum mechanically, consideration has only been given to an active detection system. Consideration has not been given to the possibility of a passive detection system. The reader who insists that such technology is impossible is being short sighted and narrow minded. We don't know what the future holds and what new ideas and/or scientific ideas may be developed. Merely because something is completely impossible for today, that doesn't necessarily mean it will never become a reality. The value of passive detection within the framework of quantum mechanics has been shown to be valid as Schrodinger's cat has been overturned.

V Is there a valid relation between energy and time?

It is only legitimate to forewarn the reader that this fifth section is even more abstract than sections II, III, and IV. The purpose of this section is to provide some commentaries on the validity of the equation $\Delta E \Delta t \geq \frac{1}{2}\hbar$.

Firstly, serious doubt should have been cast upon the validity of $\Delta E \Delta t \geq \frac{1}{2}\hbar$ by the principles outlined in this paper (especially section I). Einstein was the one who initially developed $\Delta E \Delta t \geq \frac{1}{2}\hbar$ with the intention of overturning $\Delta x \Delta p$. Namely, if both energy and time could be ascertained with precision, then so could position and momentum. (As an aside, Bohr showed that Einstein's Gedanken was flawed, and Einstein was forced to acknowledge that Bohr was right). The relation between energy and time has been extrapolated from (and predicated upon) the relation between position and momentum. If the reader agrees that both position and momentum can be simultaneously ascertained with accuracy, then that should automatically cast doubt upon the validity of $\Delta E \Delta t \geq \frac{1}{2}\hbar$.

Secondly, like Einstein, I have my own philosophical objections to the equation $\Delta E \Delta t \geq \frac{1}{2}\hbar$. Unlike Einstein's opposition to "God playing dice", my opposition lies with the laws of physics being violated merely because "special circumstances" are prevailing. Taking a very broad view of the equation for energy and time, an aspect of the equation is that if certain circumstances prevail (in this instance, time being kept sufficiently short) then the laws of physics can be broken. I personally view this as sheer and absolute nonsense. In making the transition from classical mechanics to quantum mechanics there is no question that many elements will change. As one example, classically, angular momentum can assume any value whatsoever. However, as the Stern-Gerlach experiment has shown, quantum mechanically, angular momentum is quantized. Although some elements (perhaps many elements) will change in making the transition from classical to quantum mechanics, I don't believe that the violation of conservation of energy is one of them. Merely because we don't know about it (and this, more or less is what $\Delta E \Delta t$ is stating), that doesn't mean the laws of physics can be broken. Let's draw a macroscopic analogy to illustrate the foolishness of this concept (namely, that laws of physics can be broken given the provision we are oblivious to it).

      Let's assume we have a forest that has a stream running through it, but there are no human observers anywhere near it in order to "monitor" whether the laws of physics are being broken or not. Merely because we are oblivious to it, can the reader honestly believe that rocks are floating upwards in defiance of gravity, gases are *not* in constant motion in contrast to the laws governing gases, or light which enters the stream does *not* refract? If the reader believes that such things are possible (merely because nobody is there to see it) then perhaps the reader should give up physics and become a reporter for the National Enquirer instead. In fact, an experiment can be done to test the possibility that laws of physics are being broken if there is nobody there to see it. Passive detection will be employed for this test. The floor in a room shall be constructed to detect pressure. Any increase or decrease in pressure will be registered on a computer (far away from the room) that people will monitor. Place several rocks on the floor in the room and seal the room so that nobody can look inside. If there is a decrease in pressure on the floor then we know that the rocks are floating upwards in defiance of gravity. If not, then the laws of physics are being obeyed, even if nobody is there to see it.

      Again, in making the transition from classical mechanics to quantum mechanics, many things will change. However, the law of conservation of energy is not one of them. Merely because "special circumstances" prevail (time scales on the order of $10^{-43}$ seconds) that doesn't mean that a fundamental law like conservation of energy is being broken, and a virtual photon is magically appearing out of nowhere.

      Before proceeding, it would be appropriate to clarify my position on virtual particles as what has been stated is highly conducive to misinterpretation. I am an advocate of the concept of virtual particles *given the provision that no conservation laws are being broken*. As an example, if a powerful laser beam is introduced into a closed system comprised of a vacuum, an electron-positron pair is created. However, in this instance no conservation laws are being broken as a result of the laser. In quantum field theory conservation laws *are* being broken, even if it's over an incrementally small time.

      The third area of commentary would be as follows. When I outlined an experiment to see if rocks can float upwards when nobody is looking, passive detection was relied upon. In this paper, a highly advanced passive detection system has been advocated for quantum mechanics. To put it succinctly, new technology is being advocated for the purpose of taking measurements. In a similar way, highly advanced technology may enable us to detect the virtual particles of quantum field theory (that is if they're actually there), even if those time scales are on the order of $10^{-43}$ seconds. Passive detection which assesses how something is "making its presence felt" may be the most reliable.

      In order to provide a superficial analogy for this example, I shall rely on Isaac Asimov and an analogy that he outlined. When I first started studying physics many years ago, a book by Asimov was one of the first I read. The analogy he drew in order to explain $\Delta E \Delta t$ was a chapter he called, "Behind the Teacher's Back." A teacher imposes a rule on a student. Don't stick your tongue out at me. But every time the teacher turns his back, the student sticks his tongue out at the teacher. By the time the teacher turns his back again, the student's tongue is back in his mouth. As far as the teacher is concerned, the rule is being obeyed, even though in reality, it isn't. Therefore, the correct rule is, don't let me catch you sticking your tongue out at me. Asimov then went on to state that all human rules are along these lines.

Let's reanalyze this analogy in light of new, advanced technology. Let's assume there was some device that was hand held. When the teacher turned his back to face the class, he could point it at the student's lips and assess if there was some incremental increase in saliva on the boy's lips. If there was, he would know that he stuck his tongue out at him, and realize the rule was broken. Even if the boy realized this is how the technology worked and wiped his hands across his lips to remove any excess saliva, I could still take the analogy further. Before the teacher turns to the blackboard, the technology makes a precise assessment of the boy's lips (vis-a-vis the precise epithelial cellular "structure/content" of the lip) in order to prevent the boy's ability to remove the "incriminating" evidence. If there is any significant change, the teacher knows the boy wiped his lip. If it transpires a sufficient number of times, the teacher knows the boy is sticking his tongue out at him every time he turns his back. Halliday and Resnick state it in another way. They state it's comparable to taking more money than you have out of your account but replacing it before the bank notices it was gone. However, what if they programmed their system in such a way that the *exact instant* someone overdrew on their account, that account was flagged and recorded. The next day the bank manager would have an automatic printout of patrons who engaged in this activity. The patron would then be fined (or reported to the police) for violating the rules. Whatever analogy can be drawn in order to illustrate the "validity" of $\Delta E \Delta t$, I can refute the analogy to one extent or another.

    I realize that in attempting to refute Asimov's analogy this went too far and became somewhat ridiculous. Therefore, the germane points are as follows. In regards to the two-slit electron interference pattern, Feynman has stated that if some device could be developed which could ascertain which hole the electron went through without simultaneously destroying the interference pattern, quantum mechanics would be forced to change. In regards to $\Delta E \Delta t$ a similar concept may apply. Namely, if, somehow, the virtual photons of QED could be detected when they were in the (alleged) process of mediating the electromagnetic field, then quantum mechanical theory would have to change. (The word alleged has been incorporated as I do not believe that conservation of energy can be broken for the reasons outlined under the second argument presented in this section.) I firmly believe that one day in the future I will be able to develop an advanced technology that operates on the basis of passive detection (even if the rate of progress is incremental in nature) that will enable us to detect a quantum entity without collapsing its wave function. In a similar way, it may be possible to experimentally verify whether the electromagnetic field is being mediated by virtual photons or not.

    The fourth and final area of opposition to the equation for energy and time lies in the "relationship" between energy and time. In order to expound on the meaning of this, it will be necessary to resort to extremely loose terminology. Let's take the Heisenberg uncertainty principle for position and momentum. Independent of the precepts outlined in this paper, if we keep one of these measurements incrementally small, we will get a "prize" (this is the loose terminology). For example, if we keep our measurement of position very small, we get the "prize" of an accurate measurement of momentum. It is evident that the converse is also true. Regarding the equation for energy and time, if we keep time incrementally small, we will get a "prize" of a virtual photon, gluon, etc. However what would the converse of this be? If we confine our measurement of energy to the tiniest possible amount, what's the prize? For $\Delta E \Delta t \geq \frac{1}{2}\hbar$ to be valid, there must be an intricate relation between the two. This has been the convention with position and momentum. It's superficially comparable to the Chinese concept of yin and yang. Specifically, although certain qualities may be opposite, they are in fact intricately linked with each other. One cannot exist without the other. We know what hot is because we have a frame of reference of what cold is (and

vice-versa). We know what fast is because we know what slow is. We know what tall is because we know what short is. If energy and time are genuinely related, what do we get if we keep a measurement of energy incrementally small? Even independently of Chinese philosophy, there should be a converse form of a statement if there is a genuine relation between two "quantities" that are allegedly related. Therefore, what is the converse of the statement, "by keeping a measurement of time incrementally small, the law of conservation of energy can be broken,"? In other words, how do we complete the statement, "be keeping a measurement of energy incrementally small,…"? Since there is no converse statement, how can there be a true relation between the two?

To summarize section V, there are four main points of contention against the equation $\Delta E \Delta t \geq \frac{1}{2}\hbar$.

1. Doubt should have been cast upon the validity of $\Delta x \Delta p \geq \frac{1}{2}\hbar$. Since $\Delta E \Delta t$ was predicated upon the equation for position and momentum, then the validity of the equation for energy and time should also be questioned.

2. Many things will change in making the transition from classical to quantum mechanics. I refuse to believe that the law of conservation of energy is one of them. Merely because "special circumstances" prevail that doesn't mean that conservation laws can arbitrarily be broken.

3. In a similar way that a method of detecting which hole an electron went through without destroying the interference pattern would cause quantum theory to change, some advanced technology which enabled us to detect the (alleged) photons of the electromagnetic field (or the force carriers of other aspects of quantum field theory) would also induce change.

4. For these equations to be valid, there must be an intricate relation between them. When one type of statement (as it pertains to measurement) is made, there should be a converse form of it. If a measurement of time is kept incrementally small, energy appears. What's the converse of this measurement? There is none. Is there a genuine relation between energy and time?

25